\documentclass[12pt]{article}

\usepackage{amsmath}
\usepackage{graphicx,psfrag,epsf}
\usepackage{enumerate}
\usepackage{natbib}
\usepackage{url} 

\usepackage{algorithm}
\usepackage{algorithmic}
\usepackage{array}
\usepackage{multirow}
\RequirePackage{hyperref}
\usepackage{graphicx}
\usepackage{amsmath}
\usepackage{amsthm,amsmath,natbib}
\RequirePackage[OT1]{fontenc}
\RequirePackage{amsthm,amsmath}
\usepackage[english]{babel}
\usepackage[utf8x]{inputenc}
\usepackage[T1]{fontenc}

\usepackage{amsmath}
\usepackage{graphicx}
\usepackage{enumerate}
\usepackage{natbib}
\usepackage{url} 

\newcommand{\blind}{0}

\begin{document}

\def\spacingset#1{\renewcommand{\baselinestretch}%
{#1}\small\normalsize} \spacingset{1}


{
  \title{\bf A Nonparametric Test of Dependence Based on Ensemble of Decision Trees}
  \author{Rami Mahdi\hspace{.2cm}\\
    Google}
  \maketitle
}


\bigskip
\begin{abstract}

In this paper, a robust non-parametric measure of statistical dependence, or correlation, between two random variables is presented. The proposed coefficient is a permutation-like statistic that quantifies how much the observed sample $S_n :\{(X_i , Y_i ),\ i=1…n\}$ is discriminable from the permutated sample $\hat{S}_{n×n}:\{(X_i , Y_j ),$ $\ i,j = 1…n\}$, where the two variables are independent. The extent of discriminability is determined using the predictions for the, interchangeable, leave-out sample from training an aggregate of decision trees to discriminate between the two samples without materializing the permutated sample. The proposed coefficient is computationally efficient, interpretable, invariant to monotonic transformations, and has a well-approximated distribution under independence. Empirical results show the proposed method to have a high power for detecting complex relationships from noisy data. 

\end{abstract}

\noindent%
\vfill

\newpage
\spacingset{1.5} 
\section{Introduction}

A general purpose method to detect statistical dependence, or correlation, between random variables has invaluable uses in a wide array of sciences and applications \citep{LiDX2000,DCORRCosmos2014,Mahdi2012}. Linear correlation \citep{CORR1920} is one of the oldest statistical methods that are still widely used today. Though the assumption of linearity is not always realistic, the popularity of such method stems from its ease of computation, simplicity, interpretability, and high power when the assumption of linearity is satisfied. 

Several approaches have been proposed to quantify correlation, in the general case, for more complex relationships and under less stringent assumptions. Examples of these methods are the kernel based correlation \citep{CanonCorr2004,KernelCorr2013}, copula methods \citep{CopulaCorr2012}, distance correlation \citep{DCORR2007,DCORR2009}, and discretization based mutual information (MI) \citep{MI2002} methods such as the maximal information criterion (MIC) \citep{MIC2011}.

Issues that can be lacking in some of the existing methods include: low statistical power, high computation demand, lack of intuitive interpretability, or lack of a known distribution of the coefficient under independence that would enable computing a statistical confidence. More thorough details on the pros and cons of those methods and others can be found in several studies \citep{CorrsReview2014,CorrEmprComp2018}.

\section{Contribution}

A new method is presented to measure correlation with a number of good properties:
\begin{itemize}
\item
High statistical power.
\item
Intuitive interpretability.
\item
Insensitivity to outliers.
\item
Invariance to monotonic transformations.
\item
Has a well-approximated distribution under independence. 
\item
Efficiently computable. 
\end{itemize}

In the proposed methods, the question of correlation is treated as a classification problem between the observed sample $S_n:\{(X_i,Y_i ),\ i=1…n\}$ and a virtual permutated sample $\hat{S}_{n×n}:\{(X_i,Y_j ),\ i,j=1…n\}$, where the observations along the two variables are permutated. A fast algorithm is proposed to train a Random Forest (RF) of decision trees to discriminate between the two samples without materializing the permutated sample. The predictions for the alternating leave-out sample from the RF training process is then used to determine the extent of discriminability between the two samples and, hence, whether the two variables are independent.

The proposed criterion has an interpretation similar to how a human determines correlation from a scatter plot. If the data points are scattered in a distinguishable pattern that is not expected by random when conditioned on the marginal distributions for both variables, then the two variables are deemed correlated. The proposed criterion is a quantitative measure of the same logic, in that, if we are able to build a classifier that generalizes well to discriminate the observed sample from the permutated sample, then the two variables are considered dependent or correlated.

The use of random Forests \citep{RFBreiman2001} of decision trees (RFDT) to determine discriminability is motivated by:
\begin{itemize}
\item
RFDTs are non-parametric and do not make assumptions on the data or the relationship. 
\item
RFDTs are invariant to monotonic and affine transformations.
\item
RFDTs have been empirically shown to have outstanding performance in classification problems \citep{RFEmpir2006}.
\item
RFDTs, as shown in this paper, can be trained without materializing the permutated sample. This makes RFDTs a superior choice for this specific task. 
\end{itemize}

\section{Background}

Given a bivariate sample $S_n:\{(x_i,y_i ),\ i=1…n\}$, that is randomly generated from an unknown joint probability density function (pdf) $f(X,Y)$, and marginal pdfs $f^X$ and $f^Y$ for the random variables $X$ and $Y$ respectively, $X$ and $Y$ are said to be independent if and only if:

\[
f(x,y)= f^X (x)× f^Y (y),\quad ∀x,y
\]

Since the true joint and marginal pdfs are not known, we can only rely on the observed sample, $S_n$, to determine independence. However, non-parametric learning of the pdfs from the sample $S_n$ is not trivial and the sample pdfs are typically noisy. 

\cite{HOEFFD1948B} has shown that, for continuous data, one can perform a non-parametric sample test for independence based on the distance between the sample joint cumulative distribution function (cdf), $F_n (X,Y)$, and the product of the sample marginal cdfs, $F_n^X$ and $F_n^Y$:

\[
\Delta(F)= \int  \big[ F_n (x,y) - F_n^X (x)  F_n^Y (y) \big]^2 dF_n (x,y) 
\]

Hoeffding proposed a sample estimator for this criterion and derived its distribution under independence. An alternative test, based on the sample pdfs, was later proposed by \cite{ROSENBLATT1975} and it was based on the distance criterion:

\[
∆(f)=\int\int \big[ f_n (x,y)- f_n^X (x) f_n^Y (y) \big]^2×a(x,y) \ dx dy
\]

\noindent
where $f_n$ are sample kernel density estimates and $a(x,y)$ is a weight function. The sample distribution of this criterion is not invariant to the distribution of the data and its statistical power is weaker than the criterion based on the cumulative distributions distances \citep{ROSENBLATT1975,Feuerverger1993}. An extension from \cite{ROSENBLATT1975} test was also proposed by \cite{Feuerverger1993} with fewer requirements for consistency. 

In more recent years, \cite{DCORR2007} proposed the distance correlation method (dCorr) which is based on the difference between the characteristic function of the joint pdf and the product of the characteristics functions of the marginal pdfs. 

\section{Correlation Based on a Discriminability Criterion} \label{methods_sec}

Since learning the representative properties of the generator models (i.e., pdfs) from sample data is noisy, relying on such approximations can lead to a propagated error in quantifying the difference between the generator joint distribution and the joint distribution expected under independence. In contrast, the proposed methods skip this intermediate step and, alternatively, try to learn the discriminative boundary between the two joint distributions. Such a boundary exists if and only if the two distributions are different.

\noindent\textbf{Theorem 1:} (See Appendix A for Proof)

Given two joint pdf distributions $f^A (X,Y)$ and $f^B (X,Y)$, $f^A (X,Y) \neq f^B (X,Y)$ if and only if there exists a mapping function $G:(X,Y)→ c \in \{f^{A},f^{B}\}$,\ s.t. for all bivariate random sample $(x,y)$ that is identically and randomly generated from either $f^A$ or $f^B$ with equal priors $P\big((x,y) ∼ f^A \big) = P\big((x,y) ∼ f^B \big)$, 

\[
E\bigg[ I \big(G(x,y) = c \big) \mid (x,y) ∼ c\bigg] > 0.5
\]

\noindent
where $(x,y) ∼ c$ denotes $c$ was the generator distribution of $(x,y)$ and $I(v)=1$ if $v$ is true and $0$ otherwise.

\subsection{Distribution Test:}\label{distribution_test}
Based on Theorem 1, two random variables can be tested for independence with respect to a joint distribution $f^A (X,Y)$ as follows:
\[
H_0: f^A (X,Y)=f^0 (X,Y) \iff \max_G E\big[ I\big( G(X,Y) = c\big) \mid (X,Y) ∼ c \big]=0.5
\]
\[
H_A: f^A (X,Y) \neq f^0 (X,Y) \iff \max_G E \big[ I\big(G(X,Y) = c\big) \mid (X,Y) ∼ c \big] > 0.5
\]

\noindent
where $c \in \{f^A,f^0 \}$ and $f^0 (X,Y) = f^X (X)  \times f^Y (Y)$

\subsection{Dependence Criterion Based on the Generalization of a Classifier}

When trying to determine dependence from a finite sample $S_n:{(X_i,Y_i ),\ i=1…n}$, the proposed test becomes a search for a discriminative function $G_n$ between the sample $S_n$ and the permutated sample $\hat{S}_{n×n}:{(X_i,Y_j ),\ i,j=1…n}$. The classification accuracy of $G_n$ can then be assessed using the prediction accuracy for the leave out sample. 

Bootstrap aggregation of classification models, also known as bagging, makes an ideal choice to learn $G_n$ and simultaneously assess how well it classifies unseen data thanks to the repeated subsampling and data leave-out when learning the individual classifiers. 

Given a sample of observed examples $S_n$, the output of $G_n$ for every example $(x_i,y_j )$ as a test example can be formulated as:

\begin{equation}\label{gn_formual_eq}
G_n (x_i,y_j \mid S_n )= \frac{\sum^Z_{z=1} w(x_i , y_j \mid S_n^z)×g_n^z (x_i,y_j )}{\sum^Z_{z=1} w(x_i , y_j \mid S_n^z)} 
\end{equation}

\noindent
where $Z$ is the number of learned individual classifiers,

\[
w( x_i , y_j \mid S_n^z )= \left\{\begin{array}{ll}
     1 \quad if\quad (x_i,y_i )\notin S_n^z \quad or \quad (x_j,y_j ) \notin S_n^z\\
     0 \quad otherwise
  \end{array}
  \right.
\]

\noindent
and $S_n^z$ is a random sample of $n$ examples sampled from $S_n$ with replacement, while $g_n^z$ is a function selected to minimize the empirical error of a loss function in discriminating between the observed subsample $S_n^z$, and its permutated sample $\hat{S}_{nn}^z$. The loss function can have one of many possible forms of (preferably regularized) classification error \citep{Duda2001, elts_stat_learn_09}.

Note, under independence, the value in the right hand side of equation~\eqref{gn_formual_eq} is independent from the example to be scored $(x_i , y_j)$. $G_n$ is, simply, the average of all classifiers' outputs where the example was not used in training.

\subsection{Sample Test}
Given a finite sample $S_n:{(x_i,y_i ),i=1…n}$, generated by an unknown joint pdf, $f(X,Y)$, and marginal pdfs $f^X$ and $f^Y$ for $X$ and $Y$ respectively, let $G_n$ be an aggregate of discriminative functions as in equation~\eqref{gn_formual_eq}. Then:

\[
H_0: f(X,Y) = f^X (X) \times f^Y (Y)  \iff E \big[ I(G_n (x_i , y_i \mid S_n  ) > G_n (x_j , y_h \mid S_n  )) \big]=0.5 
\]

\[
H_A: f(X,Y)\neq f^X (X) \times f^Y (Y)  \iff E\big[ I(G_n (x_i , y_i  \mid S_n )>G_n (x_j,y_h \mid S_n )) \big] > 0.5 
\]

\[
\forall i,j,h=1,2,…n,\ s.t.\   j\neq h
\]

The sample test tests whether the learned classifiers aggregated by $G_n$ generalize well to discriminate between the observed and the permutated sample by giving higher predictions to the observed test sample. The distribution difference test can be carried out using various methods including the non-parametric Mann-Whettney test \citep{MannWhitney1947}.

\subsection{Sample Coefficient of Dependence}
Given a finite sample $S_n : {(x_i,y_i ), \ i=1…n}$, let $G_n$ be an aggregate discriminative model as in equation~\eqref{gn_formual_eq}, and let $\hat{S}_m$  be a subsample from the permutated sample $\hat{S}_{n×(n-1)}:{(X_i,Y_j ),\ i\neq j}$ where $m  \leq n \times (n-1)$, a correlation based on a discriminability criterion can be expressed as:

\begin{equation}\label{rho_n_u_eq}
\rho_{nm}^U=\frac{1}{n×m} \sum_{i=1}^n \sum_{(x_j,y_h )\in \hat{S}_m} Q(G_n (x_i,y_i \mid S_n ), G_n (x_j,y_h \mid S_n )) 
\end{equation}

\noindent
where 
\[
Q(v_1,v_2 )=\left\{\begin{array}{ll}
	\  1\quad if\quad v_1 > v_2\\
    \  0\quad if\quad v_1 = v_2\\
    -1\quad if\quad v_1 < v_2
\end{array}\right.
\]

Though $\rho_{nm}^U$ is bounded between $-1$ and $1$, when $X$ and $Y$ are dependent, $G_n$ is expected to generalize well and to produce higher scores for the sample $S_n$ and, thus, $\rho_{nm}^U$ is expected to be significantly greater than zero:
\[
0<E(\rho_{nm}^U \mid H_A )  ≤ 1
\]

However, under independence, $G_n$ cannot discriminate between the two samples (theorem 1):

\[
P(G_n (x_i,y_i \mid S_n )< G_n (x_j,y_h \mid S_n,j\neq h) \mid H_0 )=P(G_n (x_i,y_i \mid S_n )> G_n (x_j,y_h \mid S_n ) \mid H_0 )=0.5
\]

As a result, $\rho_{nm}^U$ will be centered around zero
\[
E(\rho_{nm}^U |H_0 )=0
\]

And similar to the U-statistic test \citep{MannWhitney1947}, under independence and the assumptions:
\begin{itemize}
\item
A1: number of ties (equation~\eqref{rho_n_u_eq}) is small
\item
A2: $n$ and $m$ are large ($m>8$, $n>8$)
\item
A3: $G_n (x_i,y_j \mid S_n )$ are random iid withdrawals from a random variable and independent from $(x_i,y_j )$,
\end{itemize}

$\rho_{nm}^U$ has a normal distribution with 0 mean and a variance \citep{MannWhitney1947}:

\[
\sigma^2 (\rho_{nm}^U \mid H_0, A_{1,2,3})=\frac{1 + n + m}{3 \times n \times m}
\]

Though assumptions $A1$ and $A2$ are easy to satisfy, it turned out that assumption $A3$ is broken, at least when using the proposed classification method. Though $G_n (x_i,y_j \mid S_n )$ is independent from $(x_i,y_j)$, $G_n$ can produce correlated values. For example, two spatially close-by examples are likely to have more similar $G_n$ scores than would two other examples that are farther apart. This smoothness creates autocorrelations in the resulting $G_n$ values. Similar effects of such correlations on the $U$ statistic was reported in previous studies \citep{DepEffect1971,DepEffect1975,DepEffect2008}. Though such dependence in the scores can change the expected variance of $\rho_{nm}^U$, it does not change its expected value. In appendix B, we revisit this issue and show that under the assumption:

\begin{itemize}
\item
$A3^-$: $G_n (x_i , y_j \mid S_n )$ is independent from $(x_i , y_j )$ with an iid sampling violated by a weak dependency caused by localized correlations,
\end{itemize}
the variance of $\rho_{nm}^U$ can be expressed as:
\begin{equation}\label{new_null_dist_eq}
\sigma^2 (\rho_{nm}^U \mid H_0,A_{1,2,3^{-}} )) = \frac{1+n+m \times (1+K(\theta,n))}{3\times n\times m}
\end{equation}

\noindent
where $K(\theta, n)$ depends on the classification method, $\theta$, and is not sensitive to $n$. In the results section, we show empirical evidence that $K(\theta,n)$ can be well extrapolated with limited error by a constant value of 0.5. 

Note, the subset, $\hat{S}_m$, of the permutated sample is used in the test, as opposed to all of it, to reduce computation and memory requirements. Though the proposed algorithm eliminates the need to materialize the permutated sample for training, it remains necessary to represent a subset of it to evaluate the discriminative model.

\subsection{Random Forests of Second Order Partitioning Trees}

To maximize the power of detecting correlation, it is critical that the classification machine in equation~\eqref{gn_formual_eq} is capable of learning the discriminative boundary and generalizing well to test data. Moreover, learning large number of the bootstrapped individual classifiers can be computationally impractical for many classification methods, as $n$ grows large. 

Fortunately, decision trees are ideal non-parametric classification methods that can be learned fast and without, physically, materializing the permutated sample. The number of examples from the permutated sample that fall in any partition can be analytically computed. This, in turn, makes evaluating any candidate partitioning fast. 

Decision trees are usually constructed as a sequence of partitions that minimize an impurity criterion such as classification error, entropy, or $Gini$ index. Without loss of generality, in this work we use the Gini index which is a popular choice in classification problems \citep{Duda2001}. In a two class ($A$, $B$) classification problem of imbalanced data where $P_n (A) < P_n (B)$, a weighted $Gini$ impurity for a set of examples $D$ can be defined as:

\begin{equation}\label{node_split_criteria_eq}
L_{Gini} (D, \omega)=2\times \frac{| D_A |}{| D |_\omega} \times \frac{\omega \times | D_B | }{| D |_\omega}
\end{equation}

\noindent
where $|D|_\omega$ is a weighted sample size $(|D_A | + \omega\times |D_B |)$, $|D_A |$ and $|D_B |$ are the number of examples of class $A$ and class $B$ in $D$ respectively, while $\omega$ is positive coefficient. When $A$ is the observed data class and $B$ is the permutated data class, $\omega =1/n$.

Usually, the search for the next best partitioning is done by searching for the best single slice in one of the tree leaf nodes that is perpendicular to one of the variables. This univariate search, however, is not suitable for discriminating between the observed and permutated sample because they have identical marginal distributions for both variables. In fact, for the first partition, any univariate slicing will lead to no reduction in the Gini criterion. 

\begin{figure}[ht]
  \centering
  \includegraphics[scale=0.25]{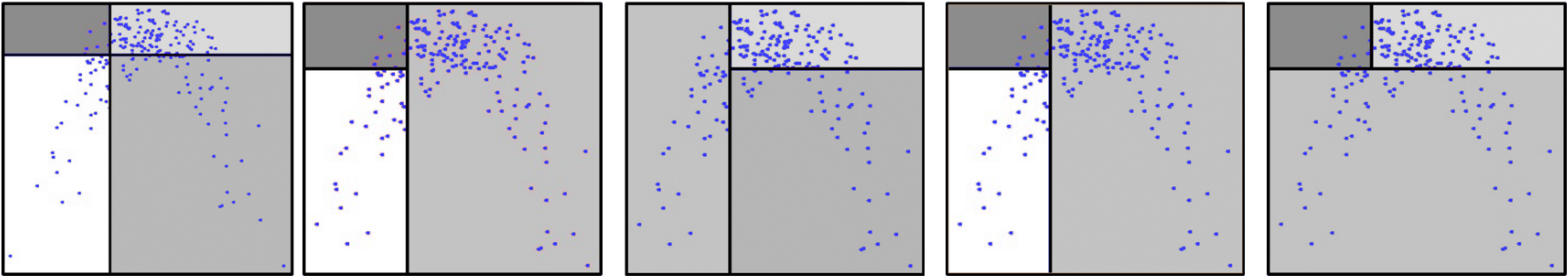}
  \caption{Within every leaf, any point within the box can be a split point. Given a split point, the box can be split in 7 different ways: horizontal, vertical, or either of the 5 shown 2-ways splits.}
  \label{fig:example1}
\end{figure}

To effectively separate the observed and permutated sample using a small size tree, a second order partitioning method is needed where partitioning along both axes, jointly, is also considered. In the proposed algorithm, every node in the tree is represented as a bounded rectangular area that is candidate for partitioning in one of seven different ways. A number of random points inside the area are selected and the seven ways of partitioning are assessed. The first two ways of partitioning are the simple univariate partitioning (vertical or horizontal) while the other five ways involve a simultaneous split along both variables as shown by the example in Figure~\ref{fig:example1}.

The criterion to select the next best partitioning is based on the reduction of the sample Gini impurity penalized by the number of resulting partitions as follows:

\begin{equation}\label{node_split_criteria_eq2}
\Delta L_{Gini} (D, \omega, \theta^k_i) = \frac{2}{|\theta_i^k|} \times
\bigg[
	| D|_\omega \times L_{Gini} (D,\omega) - \sum_{\theta^k_{i,j} \in \theta^k_{i}} \big | \langle  D\mid \theta^k_{i,j}\rangle \big|_\omega \times L_{Gini}\big (\langle D\mid \theta^k_{i,j}\rangle , \omega \big)
\bigg]
\end{equation}

\noindent
where $\theta_i^k$ is the partitioning resulting from the split point $i$ and the $k^{th}$ way of partitioning, $|\theta_i^k|$ is the number of resulting partitions (2, 3, or 4), $\theta_{i,j}^k$ is the $j^{th}$ resulting partition (sub rectangle), $\langle D|\theta_{i,j}^k\rangle$ is the set of all examples falling in $\theta_{i,j}^k$, and $|\ |_\omega$ is a weighted sample size as in \eqref{node_split_criteria_eq}.

Using the criterion in equation~\eqref{node_split_criteria_eq2}, the algorithm finds the next sub-rectangular area (tree leaf node) to split into non-overlapping 2-4 smaller rectangular areas in a way that, maximally, reduces the sum of the sample Gini impurity in the leaf nodes. This algorithm can be implemented efficiently. The first step is to convert the training sample $S_n^z :{(x_i^z,y_i^z ),\ i=1…n}$ into a rank order space, in that, we learn mapping functions $R_z^X$ and $R_z^Y$ for both variables such that:

\begin{equation}\label{x_rank_trans_eq}
R^X (v|S_n^z ) = \sum_{j=1}^n I(v \leq x_j^z)
\end{equation}
\begin{equation}\label{y_rank_trans_eq}
R^Y (v|S_n^z ) = \sum_{j=1}^n I(v \leq y_j^z)
\end{equation}

\noindent
where
\[
I (v) = 1\ if\ v\ else\ 0
\]
Using $R^X$ and $R^Y$, the training sample $S_n^z$ is transformed into the new rank space as $T_n^z:{(r_i^{xz},r_i^{yz} )= (R^X (x_i^z \mid S_n^z ),\ R^Y (y_i^z \mid S_n^z )),\ i=1…n}$, and the permutated virtual sample becomes $\hat{T}_{nn}^z:{(r_i^{xz},r_j^{yz} )\ ,i,j=1…n}$. As a result of this transformation, given any rectangular area, in the new space, that spans $(r_i^{xz},r_t^{xz} \rbrack$ on the $x$-rank-axis and $(r_j^{yz},r_l^{yz} \rbrack$ on the $y$-rank-axis, the number of permutated examples that fall within this area can be analytically and exactly computed as:

\begin{equation}
\label{premu_examples_num_eq}
\bigg|  \big \langle\hat{T}_{nn}^z \mid \big( r_i^{xz}, r_t^{xz} \big],\ \big( r_j^{yz},r_l^{yz}  \big] \big \rangle  \bigg| 
= ( r_t^{xz} - r_i^{xz} )×( r_l^{yz} - r_j^{yz} )
\end{equation}

This preprocessing step eliminates the need to represent/materialize the permutated sample and reduces the computationally complexity dramatically. Algorithm~\ref{alg1} outlines the proposed tree learning algorithm. 

\begin{algorithm} 
\caption{: Decision Tree Learner}
\label{alg1}
\begin{algorithmic} [1]
\REQUIRE $S_n^z:{(x_i^z,y_i^z ),i=1…n}$
\REQUIRE maxLeafCount, splitTrialCount, leafMinWidth
\REQUIRE $\omega = 1/n$

\STATE Let $R_z^X$, $R_z^Y$ be the rank mapping functions as in \eqref{x_rank_trans_eq} and \eqref{y_rank_trans_eq}.
\STATE Let $T_n^z$ be the rank transformation of $S_n^z$: $T_n^z:{(r_i^{xz},r_i^yz )= (R^X (x_i^z |S_n^z ),R^Y (y_i^z |S_n^z )),i=1…n}$

\STATE root = TreeNode<sample=$T_n^z$ ,  xRange=(0,n] , yRange=(0,n]>
\STATE LeafNodes = \{ root \}  

\REPEAT
\STATE $\langle leaf_t , \theta^k_i \rangle =  arg \displaystyle\max_{leaf_t \in LNodes ,\theta_i^k} \Delta L_{Gini} (\langle D\mid leaf_t \rangle, \omega, \theta_i^k)  $
\STATE $newLeafNodes  \gets$ split the tree node $leaf_t$ according to $\theta_i^k$
\STATE $LeafNodes =  \big( LeafNodes - \{ leaf_t \} \big)\ \cup \   newLeafNodes$
\STATE $leaf_t .childern = newLeafNodes$
\UNTIL{$|LeafNodes| \geq MaxLeafCount$ }
\STATE return $\langle root,  R_z^X, R_z^Y \rangle$

\textbf{\%Implementation details:} To find a good way to split a leaf node
\\ \% \quad A number (trialCount) of random points are selected within the node rectangle
\\ \% \quad For every candid split point:
\\ \% \qquad splitting into 4 sub rectangles is evaluated first
\\ \% \qquad\quad Number of examples from both classes in every sub rectangle is counted
\\ \% \qquad\qquad Number of permutated examples is computed using equation~\eqref{premu_examples_num_eq}
\\ \% \qquad\quad Computing $\Delta L_{Gini}$ for the seven ways of partitioning is now straightforward
\\ \% \quad Largest $\Delta L_{Gini}$ is computed only once for every leaf and stored
\\ \% Any partitioning that leads to a node of width less than leafMindWidth is discarded

\end{algorithmic}
\end{algorithm}

\begin{figure} 
  \centering
  \includegraphics[scale=0.37]{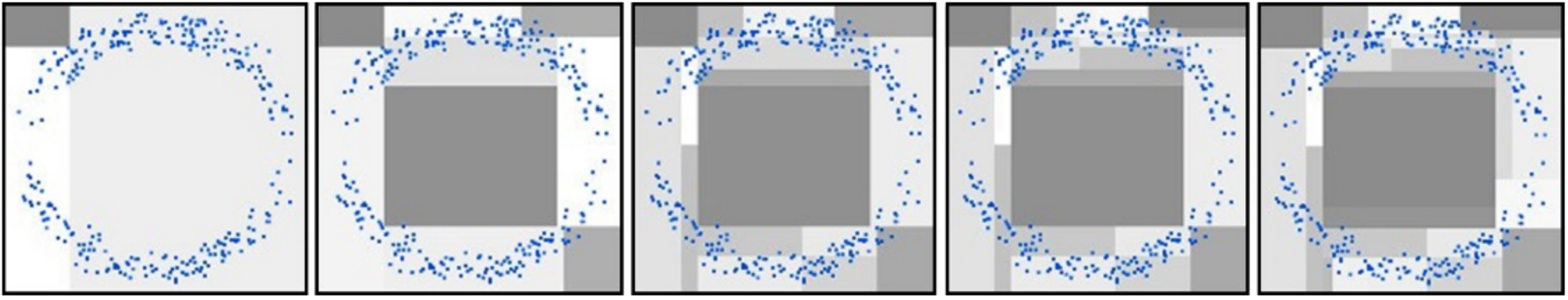}
  \caption{Example of progressive partitioning, left to right, by algorithm~\ref{alg1} to minimize the Gini impurity in a single tree. The data is based on a circle relationship with uniform noise.}
  \label{fig:example2}
\end{figure}

Once the tree is constructed using algorithm~\ref{alg1}, every leaf of the tree is assigned a label as follows:

\begin{equation}\label{leaf_label_eq}
Label(leaf_t \mid S_n^z )= \frac{|\langle D_A \mid leaf_t \rangle |}{|\langle D \mid leaf_t \rangle |_\omega}
\end{equation}

\noindent
where $|\langle D_A \mid leaf_t \rangle |$ is the number of training observed examples that fall within the boundary of $leaf_t$, while $|\langle D \mid leaf_t \rangle |_\omega$ is the number of all training examples (observed and permutated) weighted as in \eqref{node_split_criteria_eq}. The label for every leaf ranges from 0 to 1 and it represents the relative probability density between the two classes. Also, when a test example $(x_i,y_j )$ falls within the boundary of the leaf $t$, it is assigned a prediction score that equals the leaf label:

\[
g_n^z (x_i,y_j \mid (x_i,y_j ) \sim_R leaf_t )=Label(leaf_t \mid S_n^z )
\]

\noindent
where $(x_i,y_j ) \sim_R leaf_t$ denotes that the example $(x_i,y_j )$ falls within the boundary of the $leaf_t$ after rank transformation. Figure~\ref{fig:example2} shows an example of the progressive partitioning produced by algorithm~\ref{alg1}. 

\begin{figure} 
  \centering
  \includegraphics[scale=0.223]{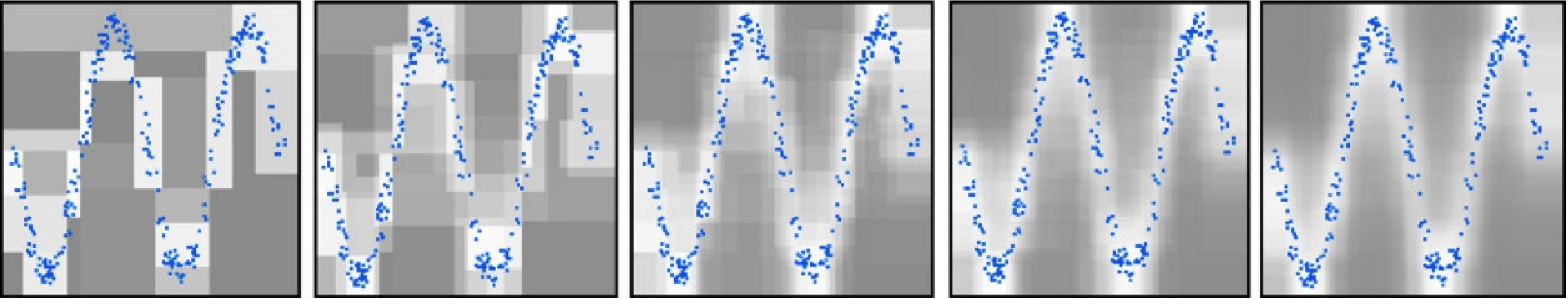}
  \caption{Learning classification boundary as an aggregate of decision trees trained to reduce leaves impurity on bootstrapped sampled data. Each tree has 23 leaves. Numbers of trees, left to right, are 1, 3, 9, 27, and 81.}
  \label{fig:example3A}
\end{figure}

Once the training of every tree is complete, it is used to score/classify examples that were not used in training. Algorithm~\ref{alg2} outlines the bootstrapping process to build multiple trees and compute the coefficient of correlation, while Figure~\ref{fig:example3A} shows examples of the resulting bootstrapped classifiers.

\begin{algorithm} 
\caption{: compute uCorr}
\label{alg2}
\begin{algorithmic} [1]
\REQUIRE $S_n:{(x_i,y_i ),\ i=1…n}$
\REQUIRE MaxTreeCount

\STATE $\hat{S}_m \Leftarrow$ sample without
replacement from $\hat{S}_{n\times n-1}:{(x_i,y_j ),\ i,j=1…n}$
\STATE $z \Leftarrow 0$
\REPEAT
\STATE $z \leftarrow z + 1$
\STATE $S_n^z \leftarrow$ sample $n$ examples from $S_n$ with replacement.
\STATE $\langle root_z , R_z^X, R_z^Y \rangle = Algorithm1 (S_n^z)$
\STATE Using $\langle root_z , R_z^X, R_z^Y \rangle$, score the examples in $S_n$ and $\hat{S}_m$ that were not used in training $root_z$
\UNTIL{$z = MaxTreeCount$}

\STATE Use the average of scores for examples in $S_n$ and $\hat{S}_m$, compute $\rho_{nm}^U$ using equation~\eqref{rho_n_u_eq}
\STATE return $\rho_{nm}^U$

\end{algorithmic}
\end{algorithm}

\subsection{Further Improvements}

For decision trees to generalize well in classifying test examples, it is critical to use a good set of training parameters including the size of the tree (number of leaves), and the minimum allowed width of every leaf. Those parameters can be selected by training a limited number of trees to assess the most appropriate values. Nonetheless, empirical results of a greedy search for those parameters did not yield a significant improvement over just setting them to reasonable values for all cases. In all experiments shown in figure~\ref{fig:main_result} and table~\ref{tbl:tbl1}, setting the number of leaves to $\sqrt[]{n}$ and the minimum width per leaf to $0.03 \times n$ was found to work well for all cases.
 
\begin{figure} 
  \centering
  \includegraphics[scale=0.24]{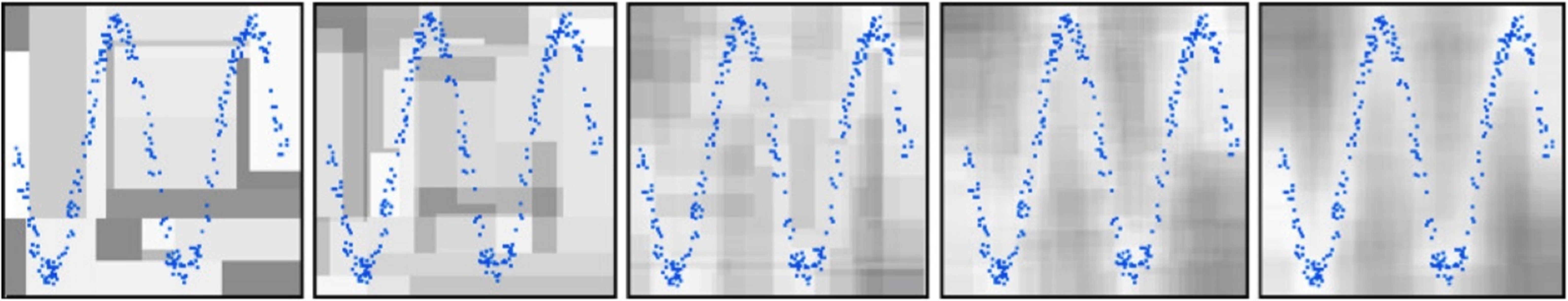}
  \caption{Learning classification boundary as an aggregate of decision trees trained by semi-random splits on bootstrapped sampled data. Each tree has 23 leaves. Numbers of trees, left to right, are 1, 3, 9, 27, and 81.}
  \label{fig:example3B}
\end{figure}

Moreover, in half of the trees, the partitioning criteria in equation~\eqref{node_split_criteria_eq2} is replaced by a semi-random selection criteria:

\[
\Delta L_{rand} (D,\omega,\theta_i^k )=\frac{2}{|\theta_i^k|} \times \gamma \times \sqrt[]{|D|_\omega}
\]
\noindent
where $\gamma \in [0,1]$ is a uniform random variable. This criteria is not only faster to compute, but it also causes the algorithm to make partitions in places where the first criteria may consistently ignore. Though the partitioning is random, the labels assigned to the leaves (equation~\eqref{leaf_label_eq}) are not arbitrary. However we partition the data, those labels remain representative of the relative probability density between the two samples. Also, it has been shown \citep{RandTrees2003} that arbitrary partitioning in bootstrapped decision trees work as well, and sometimes improves upon, impurity reduction methods. Figure~\ref{fig:example3B} shows examples of the resulting bootstrapped classifiers based on the proposed random splits. 

\section{Computational Complexity}

The computational cost of evaluating a single candidate split of a node $t$ that encloses $n_t$ of the observed examples is of the order $O(n_t)$. This is because we need only to count the observed samples that fall in each of the four sub-areas, while counting the number of permutated examples (equation~\eqref{premu_examples_num_eq}) is of a constant order and, thus, is negligible. 

Since the number of examples enclosed in all nodes at any layer in the tree is always $n$, the total complexity of evaluating one candidate split in each node in one layer of the tree is $O(n)$. In the worst case, the number of layers in the tree is the same as the number of the leaf nodes and, thus, the worst total complexity of a single tree training is $O( n \times \text{number of leaves} \times \text{number of split trials})$. In addition, to train $Z$ number of trees, the total training complexity is $O( Z \times n \times \text{number of leaves} \times \text{number of split trials})$. 

Since the number of trees, the number of split trials, and the number of leaves are relatively small in value, the over all complexity shrinks to $O(\text{constant} \times n)$. In all experiments reported, the number of trees was 100, one third of which were partitioned randomly with negligible computation. In addition, the number of split trials was set to 10, while the number of leaf nodes was at most 64. This brings down the overall complexity to $O( 64,000 \times n)$. 

The most important take away of this complexity result is that the computationally complexity of computing the proposed coefficient of correlation grows linearly in $n$ which means it is scalable and suitable to use for large $n$ data cases.

\section{Experiments \& Results} \label{results_sec}

In this section, the proposed coefficients is compared to five other correlation methods: linear Pearson correlation (pCorr), distance correlation (dCorr), maximal information criterion (MIC), Hoeffiding distance (HeoffD), and the randomized dependence coefficient (RDC). 

The first experiment is a case study of 5 simulated examples that shows the effect of increasing noise on the value of the 6 coefficients and how they decline as the noise increases. Table~\ref{tbl:tbl1} show the value of the 6 coefficients for three levels of noises. We see that, for all cases, uCorr, similar to RDC and MIC show large values (>0.5) when noise was limited and continued to decline as noise increased while remaining higher than the range of values under $H_0$. In contrast, dCorr and HeofD had low coefficient values even at low noise levels for some of the relations.

\renewcommand{\arraystretch}{0.61}
\begin{table}
  \centering
    \caption{A simulation of five relationships at 3 levels of noise ($L_1$, $L_2$, and $L_3$) with sample size of 300 each. The coefficient values for each relation at each levels noise are listed. The mean and standard deviation of the coefficient when the data is permutated to emulate the null hypothesis are listed under $H_0$ column.}
  \label{tbl:tbl1}
  \begin{tabular}{ | c | c | c | c |m{2.3cm}| m{7.7cm} | }
    \hline
    noise & L1 & L2 & L3 & \quad H0 & \quad L1 \qquad\qquad\qquad  L2 \qquad\qquad\qquad L3 \\ \hline
    uCorr & 0.93 & 0.81 &  0.48 & 0.001 $\pm$ 0.05  
    &
    
    \multirow{6}{*}{
	    \hspace{-4mm}
    	\begin{minipage}{.1\textwidth}    	
      	\includegraphics[scale=0.325]{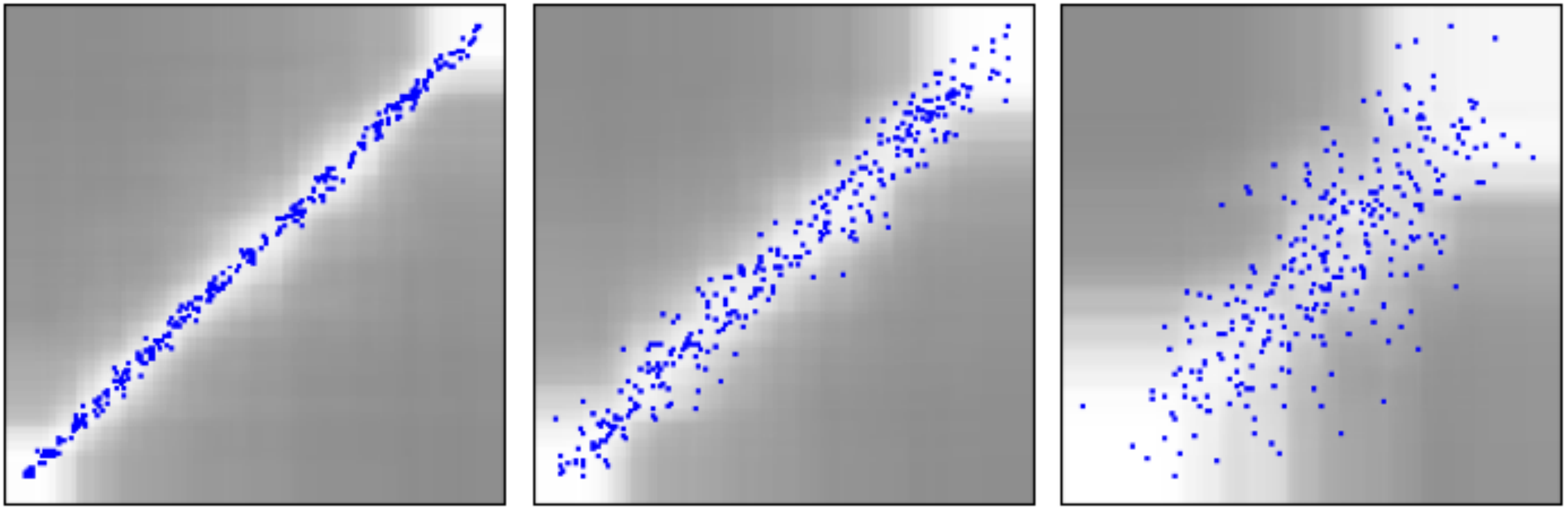}
    	\end{minipage}
    }
    \\ \cline{1-5}
    pCorr & 1 & 0.98 & 0.77 & \ 0.0 {\small $\pm$ 0.6} & \\ \cline{1-5}
    dcorr & 1 & 0.95 & 0.56 & 0.01 {\small $\pm$ 0.005} & \\ \cline{1-5}
    RDC & 1 & 0.98 & 0.78 & 0.18 {\small $\pm$ 0.04} & \\ \cline{1-5}
   	HoefD & 0.91 & 0.71 & 0.25 & 0.0 {\small $\pm$ 0.003} & \\ \cline{1-5}
   	MIC & 1 & 1 & 0.57 & 0.19 {\small $\pm$ 0.02}  & \\ 
    \hline \hline
    
    uCorr & 0.81 & 0.56 & 0.15 & \ 0.0 {\small $\pm$ 0.04}
    &
    \multirow{6}{*}{
    	\hspace{-4mm}
        \begin{minipage}{.1\textwidth}
      	\includegraphics[scale=0.325]{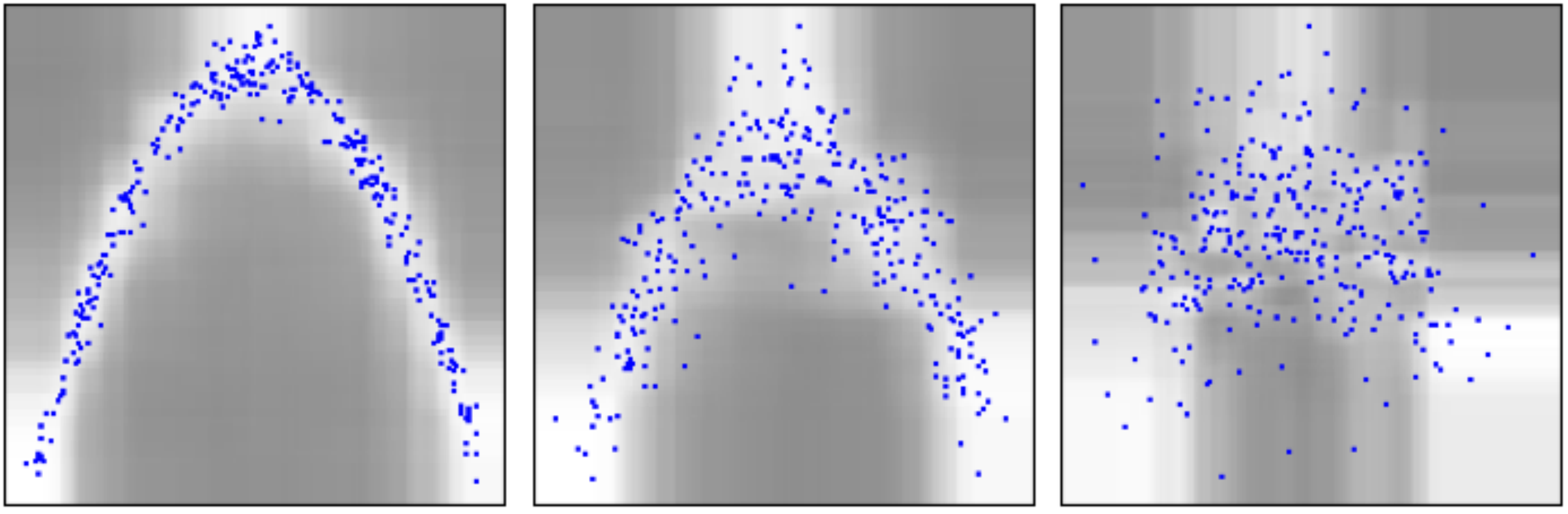}
    	\end{minipage}
    }
    \\ \cline{1-5}
      pCorr & 0.02 & 0.02 & 0.03 & 0.01 {\small $\pm$ 0.06} & \\ \cline{1-5}
      dcorr & 0.24 & 0.19 & 0.03 & 0.01 {\small $\pm$ 0} & \\ \cline{1-5}
      RDC & 0.99 & 0.87 & 0.37 & 0.18 {\small $\pm$ 0.04} & \\ \cline{1-5}
      HoefD & 0.18 & 0.09 & 0.01 & \ 0.0 {\small $\pm$ 0} & \\ \cline{1-5}
      MIC & 1 & 0.65 & 0.24 & 0.19 {\small $\pm$ 0.02} & \\
    \hline \hline

	uCorr & 0.8 & 0.64 & 0.18 & 0.002 {\small $\pm$ 0.05}
    &
    \multirow{6}{*}{
    	\hspace{-4mm}
    	\begin{minipage}{.1\textwidth}
      	\includegraphics[scale=0.325]{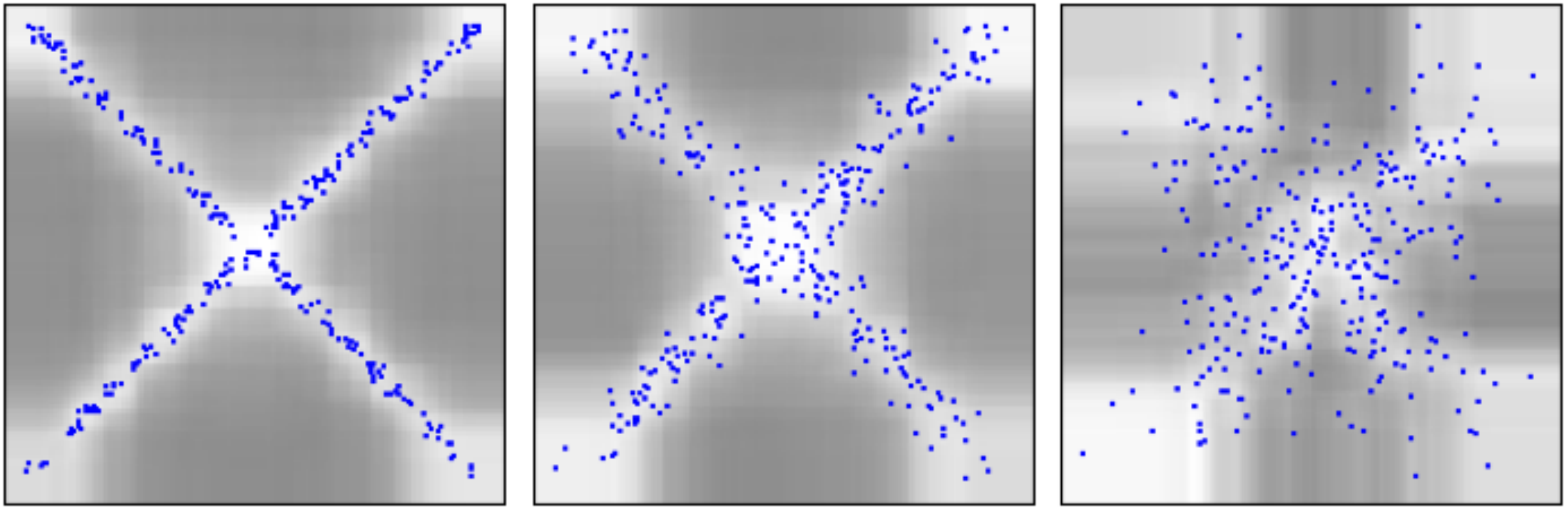}
    	\end{minipage}
    }
    \\ \cline{1-5}
    
    pCorr & 0.05 & 0.06 & 0.06 & \ 0.0 {\small $\pm$ 0.055} & \\ \cline{1-5}
    dcorr & 0.07 & 0.06 & 0.02 & 0.01 {\small $\pm$ 0.004} & \\ \cline{1-5}
    RDC & 0.99 & 0.93 & 0.43 & 0.17 {\small $\pm$ 0.04} & \\ \cline{1-5}
    HoefD & 0.05 & 0.03 & 0 & \ 0.0 {\small $\pm$ 0.003} & \\ \cline{1-5}
    MIC & 0.6 & 0.49 & 0.19 & 0.18 {\small $\pm$ 0.02} & \\
    \hline \hline

	uCorr & 0.77 & 0.54 & 0.13 & \ 0.0 {\small $\pm$ 0.04}
    &
    \multirow{6}{*}{
    	\hspace{-4mm}
    	\begin{minipage}{.1\textwidth}
      	\includegraphics[scale=0.325]{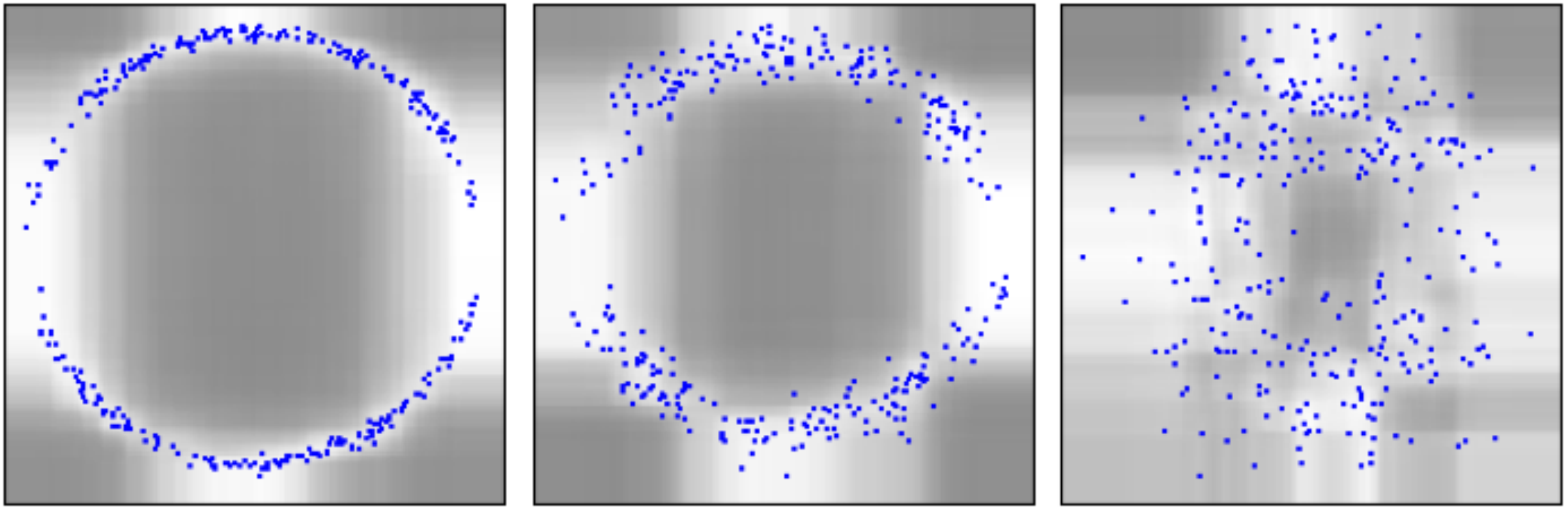}
    	\end{minipage}
    }
    \\ \cline{1-5}
    pCorr & 0.01 & 0.01 & 0.01 & 0.001 {\small $\pm$ 0.06} & \\ \cline{1-5}
    dcorr & 0.02 & 0.02 & 0.01 & 0.01 {\small $\pm$ 0.006} & \\ \cline{1-5}
    RDC & 0.95 & 0.86 & 0.35 & 0.18 {\small $\pm$ 0.04} & \\ \cline{1-5}
    HoefD & 0.05 & 0.02 & 0 & \ 0.0 {\small $\pm$ 0.001} & \\ \cline{1-5}
    MIC & 0.6 & 0.44 & 0.22 & 0.18 {\small $\pm$ 0.03} & \\
    \hline \hline

	uCorr & 0.77 & 0.56 & 0.17 & 0.001 {\small $\pm$ 0.05}
    &
    \multirow{6}{*}{
    	\hspace{-4mm}
    	\begin{minipage}{.1\textwidth}
      	\includegraphics[scale=0.325]{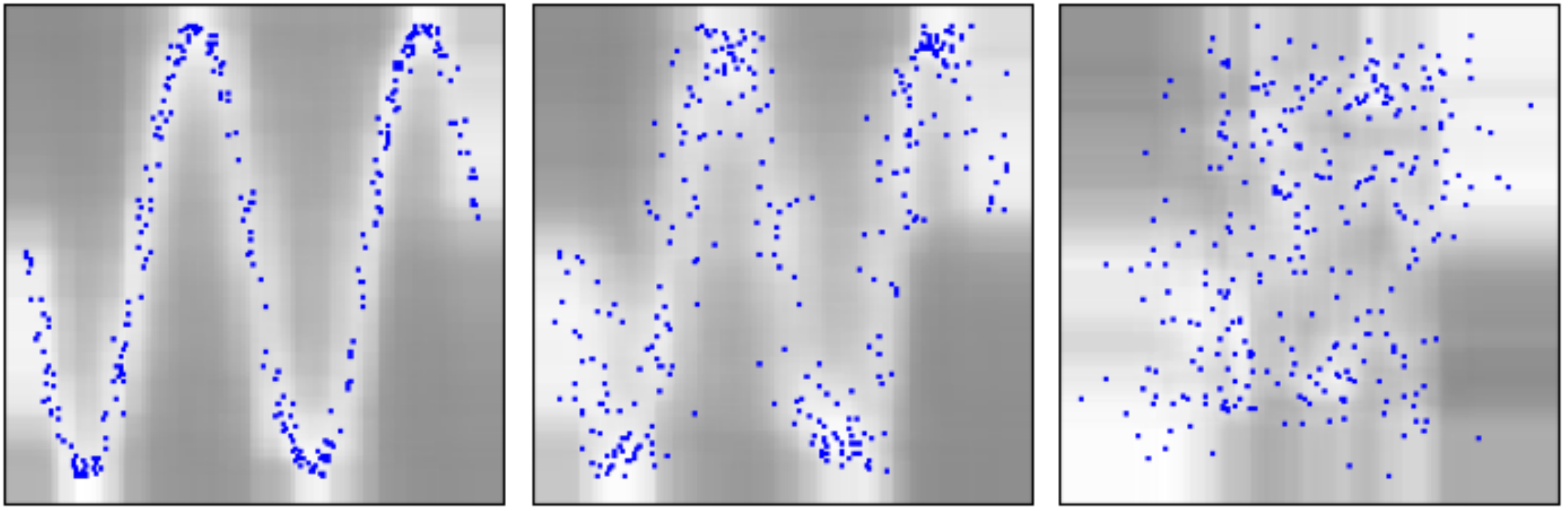}
    	\end{minipage}
    }
    \\ \cline{1-5}
    pCorr & 0.43 & 0.41 & 0.29 & 0.0 {\small $\pm$ 0.05} & \\ \cline{1-5}
    dcorr & 0.21 & 0.18 & 0.08 & 0.02 {\small $\pm$ 0.005} & \\ \cline{1-5}
    RDC & 0.69 & 0.55 & 0.32 & 0.18 {\small $\pm$ 0.05} & \\ \cline{1-5}
    HoefD & 0.11 & 0.07 & 0.02 & 0.001 {\small $\pm$ 0.004} & \\ \cline{1-5}
    MIC & 0.99 & 0.8 & 0.27 & 0.18 {\small $\pm$ 0.02} & \\
    \hline    
  \end{tabular}
\end{table}

In the second experiment, we quantify the statistical power of the 6 coefficients on 8 relationships, linear and non-linear. For each case, and each noise level we generate a sample of $n=400$ and repeat the experiment 5,000 times with different randomization seed. To simplify the comparison and make it fair, for all coefficients, the distribution of each coefficient under $H_0$ is computed empirically by permutating the data and computing the coefficients. Power is then measured as the percentage of the times the coefficient under $H_A$ is larger than the 95\% quantile of the coefficient values under $H_0$.

\begin{figure} 
  \centering
  \includegraphics[scale=0.3]{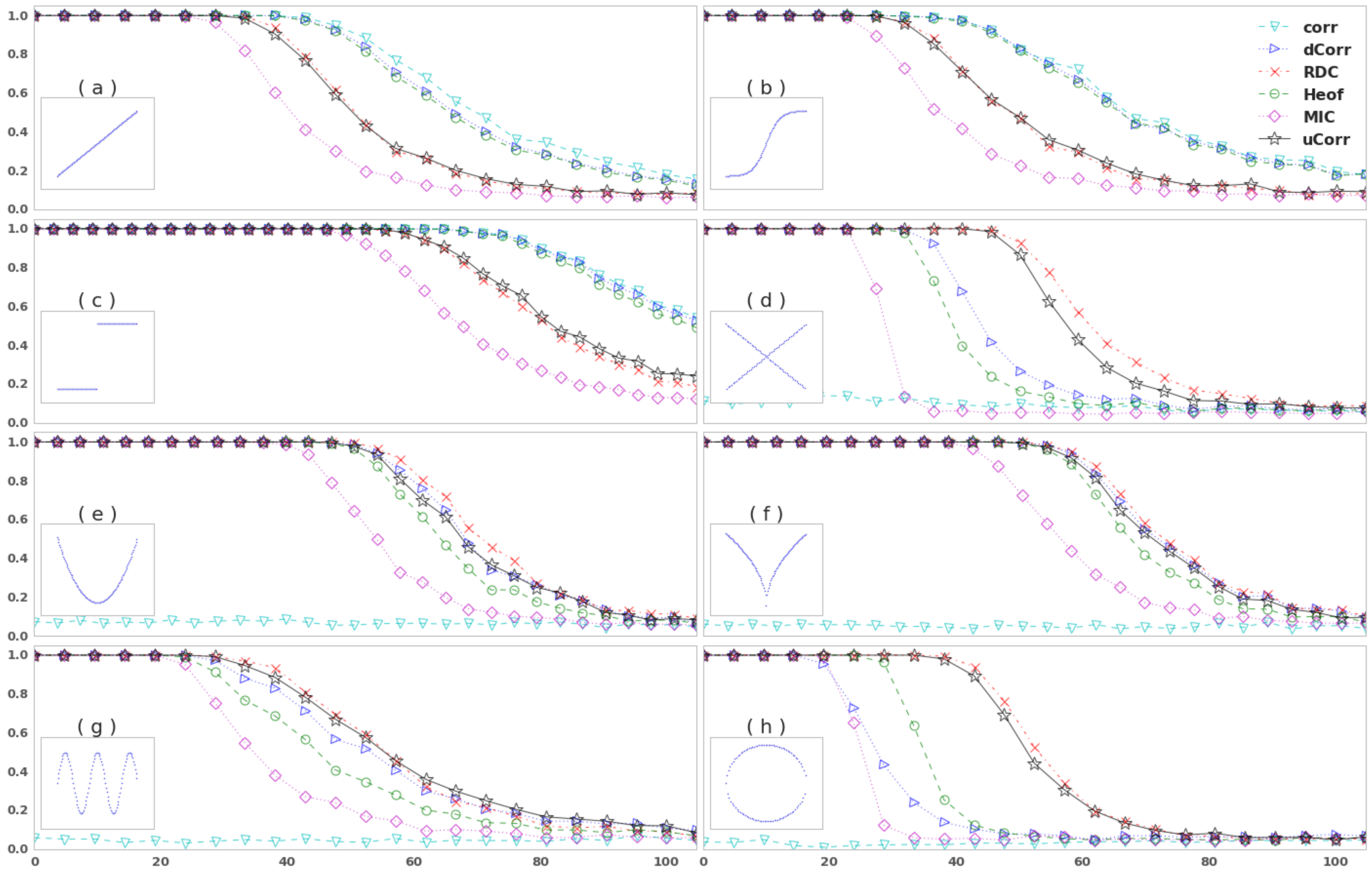}
  \caption{Statistical power of each method in detecting dependence under various relationships and increasing level of noise. Y-axis is the fraction of times dependence was correctly detected. X-axis is the level of noise, where the actual scale in each case was mapped to the range [0-100] for the simplicity of plotting.}
  \label{fig:main_result}
\end{figure}

Figure~\ref{fig:main_result} shows the statistical power of each method as noise increases on each relationship. We see that in relations a), b), and c) which have significant linear relationship component, that uCorr similar to RDC has better power than MIC, but weaker than pCorr, HeofD, and dCorr. Those later 3 methods benefit when the linearity assumption holds. However, in the other five relations that were mostly non-linear (d, e, f, g, and h), uCorr had a power performance competitive to the best performance in all cases. 

In the $3rd$ experiment, we validate the derived distribution of uCorr under independence. A random sample of two variables is generated for samples sizes of $n = 200$ and $n = 1,000$. For each $n$, the experiment is repeated 100,000 times and an empirical distribution is plotted and compared against the expected distribution, for values of $m=2,000$ and 10,0000, equation~\eqref{new_null_dist_eq}.

\begin{figure} 
  \centering
  \includegraphics[scale=0.3]{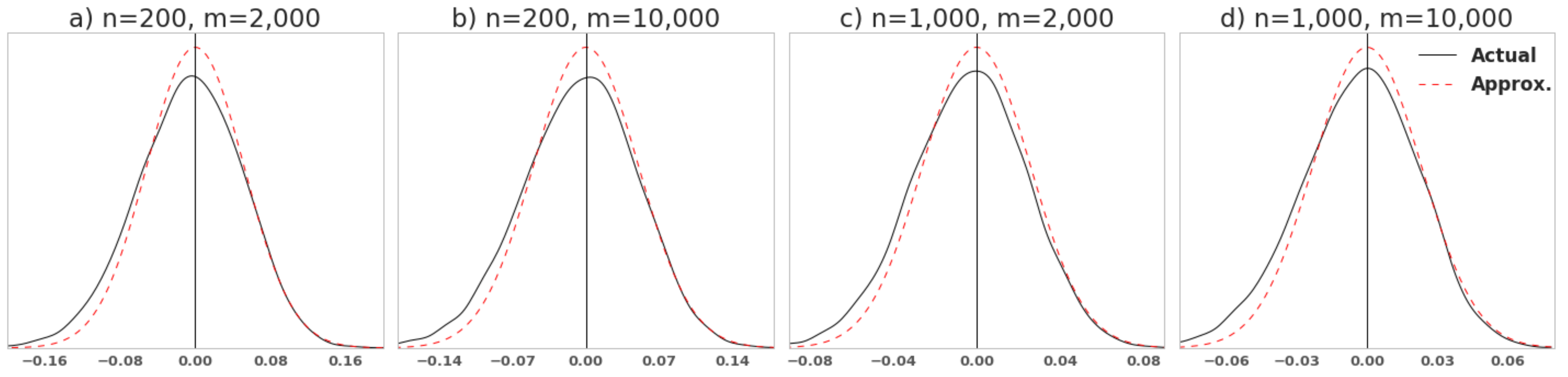}
  \caption{Distribution of uCorr under independence for varying $n$ and $m$. Solid line is the actual empirical distribution while dashed line is the analytically derived/approximated distribution. In all cases, a constant value of 0.5 for the bias term $K(n,\theta )$ is found to provide a good approximation for the actual distribution.}
  \label{fig:null_dist_comparison}
\end{figure}

From figure~\ref{fig:null_dist_comparison}, we see that the null distribution given by equation~\eqref{new_null_dist_eq} is a good approximation. We also see that the shape of the empirical distribution is not exactly symmetric around zero. One explanation could be that the effect of smoothness, represented by the term $K(\theta,n)$ in equation~\eqref{new_null_dist_eq}, is not constant for all cases. Nonetheless, the approximation seems accurate for the right most part of the curve and could serve well for approximating p-values. For higher accuracy of p-values, the user can always generate an empirical null distribution by permutation. 

\section{Conclusion}

The proposed methods in this paper redefine the statistical dependence between two variables as a classification problem between what is observed and what is expected under independence. A practical method is presented to compute a discriminability criterion based on an aggregate of decision trees. Analysis of statistical power based on simulated data shows the proposed approach to be superior to commonly used methods and competitive to the best of the competing methods. In addition, the derived approximated null distribution of the proposed coefficient is shown to be fairly accurate for different values of $n$. This can simplify computing p-values for real applications. 

Finally, some of the open questions that are not addressed in this paper include:
\begin{itemize}
\item How other classification methods would compare to decision trees for the same problem. 
\item How to extend the proposed approach to compute conditional dependence.
\end{itemize}

These are valuable questions to answer and should be investigated in future studies. 

\section{Appendix}

\subsection{Part A: Theorem 1 Proof:}

\textbf{Case 1:}

When $f^A (X,Y)=f^B (X,Y)$ and the priors are equal $P(f^A ) = P(f^B )=0.5$, for any observation $(x,y)$, $P(f^A \mid (x,y ))= P(f^B \mid (x,y ))=0.5$. With the absence of any other knowledge, any mapping function $G:(X,Y)→ c \in \{f^{A},f^{B}\}$ can only be right half of the times on average: 

\[
E \bigg[ I\big( G(x ,y )=c \big) \mid\ (x , y ) ∼ c \bigg] =0.5
\]

\noindent
where $(x , y ) ∼ c $ means $c \in \{f^{A},f^{B}\} $ was the generator distribution of the observed example $(x ,y )$.

\noindent
\textbf{Case 2:}

When $f^A (X,Y) \neq f^B (X,Y)$, and the priors are equal $P(f^A ) = P(f^B )=0.5$, for any observation $(x ,y )$, the following is self-evident by definition:
\[
P(f^A \mid (x,y)) > 0.5 > P(f^B |(x,y)) \iff f^A (x , y)>f^B (x,y)
\]
\[
P(f^A |(x,y )) < 0.5 < P(f^B |(x,y )) \iff f^A (x,y ) < f^B (x,y )
\]
\[
P(f^A |(x,y)) = 0.5 = P(f^B |(x,y))  \iff f^A (x,y ) = f^B (x,y )
\]
\noindent
An ideal mapping function $G^{*}$ can be then defined as:
\[
G^{*}(x,y ) = arg \displaystyle\max_{c \in \{ f^A , f^B \} } c ( x , y )
\]

\noindent
As a result, when $f^A (x,y ) \neq f^B (x,y )$ , $P(G^{*}(x,y ) = c \mid (x,y ) ∼ c )>0.5$ , while when $f^A (x,y ) = f^B (x,y )$, $G^{*}$ will be an arbitrary function similar to case 1 and $P(G^{*}(x , y )= c \mid (x , y ) ∼ c )=0.5$.

\noindent
Since $f^A (X,Y) \neq f^B (X,Y)$, it follows $P(f^A (x,y ) \neq f^B (x,y ))>0$, and thus $G^{*}$ is expected to map the example to the generator distribution more than half of the times:
\[
E \bigg[ I\big( G^{*}(x ,y )=c \big) \mid\ (x , y ) ∼ c \bigg] > 0.5
\]

\noindent
Based on Case 1 and Cases 2,

\[
f^A (X, Y) \neq f^B (X, Y)\ iff\  \exists G:(X, Y) \rightarrow c \in \{ f^A , f^B \},\ s.t. 
\]

\[
E\bigg[ I \big( G ( x , y ) = c \big) \mid ( x , y ) ∼ c \bigg] > 0.5
\]

\subsection{Part B: Distribution Under Independence:}
The distribution of $\rho_{nm}^U$ under independence, equation~\eqref{new_null_dist_eq}, can be derived from the distribution of a similar, but more simplified variable defined as:
\[
\hat{\rho}_{nm}^U = \displaystyle\sum_{i = 1}^{n} \displaystyle\sum_{j = 1}^{m} I_{ij} 
\]

\noindent
where

\[
I_{ij} = \left\{\begin{array}{ll}
     1 \quad if\quad G_i > G_j \\
     0 \quad if\quad G_i < G_j
  \end{array}
  \right.
\]

\noindent
and $G_i$ and $G_j$ are two random variables representing the output of the classifier for the observed example $i$ and the virtual examples $j$ respectively. Under independence and either assumption $A3$ or $A3^-$ (section~\ref{methods_sec}), 
\[
P(I_{ij} = 1) = P(I_{ij} = 0) = 0.5
\]

Under assumption $A3$, $E(I_{i'j'} )$ is independent from $I_{ij} $ for all $i\neq i'$ and $j\neq j'$, while under $A3^-$, $E(I_{ij} )$ is correlated within some sets of examples, based on their closeness in the 2D rank space. For example, given $G_i$ is large for an observed example $x_i$, that by definition implies there is at least another observed example close to $x_i$ in the 2D space, and that example will also have a high $G$ score dictated by its closeness to $x_i$.

As a consequence of $A3^-$, correlation happens among close-by examples. Also, it is reasonable to expect the size of those neighborhoods, where correlation is significant, to not grow fast with $n$ since we are merely modeling a stochastic clustering of examples under independence.

Taking $A3^-$ in consideration, the variance of $\hat{\rho}_{nm}^U$ can be derived as:
\[
Var(\hat{\rho}_{nm}^U )=Var \bigg( \sum_{i=1}^n \sum_{j=1}^m I_{ij} \bigg)
=
\sum_{i=1}^n \sum_{j=1}^m \sum_{i' =1}^n \sum_{j' =1}^m C_{ij\_i'j'}
\]

\[
  \begin{aligned}
= \sum_{i=1}^n \sum_{j=1}^m \bigg[ 
	C_{ij\_ij} +
    \sum_{i^{''} \notin T_n (i)} C_{ij\_i^{''} j} + 
    \sum_{i^{'} \in T^{-}_{n} (i)} C_{ij\_i^{'} j} + 
    \sum_{j^{''} \notin T_m (j)} C_{ij\_ij^{''}} +
    \sum_{j^{'} \in T^{-}_{m}(j)} C_{ij\_ij^{'}} + \\
    \sum_{i^{''} \notin T_n (i)} \sum_{j^{''} \notin T_m (j)} C_{ij\_i^{''} j^{''}} +
    \sum_{i^{'} \in T^{-}_{n} (i)} \sum_{j^{'} \in T^{-}_{m}(j)} C_{ij\_i^{'} j^{'}} +
    \sum_{i^{''} \notin T_n (i)} \sum_{j^{'} \in T^{-}_{m} (j)} C_{ij\_i^{''} j^{'}} + \\
    \sum_{i^{'} \in T^{-}_{n} (i)} \sum_{j^{''} \notin T_m (j)} C_{ij\_i^{'} j^{''}} 
	\bigg] 
  \end{aligned}
\]

\noindent
where $C_{ij\_ i'j'} = Covariance( I_{ij},\ I_{i'j'} )$, $T_n (i)$ is the set of the example $i$ and the examples surrounding it, where localized-correlation of the model output is significant enough not to ignore, while $T^{-}_{n} (i) = T_{n}(i) - \{example \ i\}$. If those neighborhoods exist under independence, their expected size can, only, be a function of either the model parameters, $\theta$, the sample size $n$, or both: 

\[
	E[ | T_n (i) | ] = 1 + E[ | T^{-}_{n}(i) | ] = 1 + a ( n , \theta)
\]

\noindent
and similarly,
\[
	E\big[ | T_m (j) | \big] = 1+ E\big[ | T^{-}_{m} (j) | \big] \approx 1 + a(n,\theta) \times \frac{m}{n}
\]

\noindent
Also, from applying permutation rules to independent variables we get:
\[
Cov(I_{ij} , I_{ij}  ) = \frac{1}{4}
\]
and
\[
Cov(I_{ij} , I_{i^{''}j^{''}} ) = 0,\ \ Cov(I_{ij} , I_{i^{''}j} ) = Cov(I_{ij} , I_{ij^{''}} ) = \frac{1}{12},\ 
\forall i^{''} \notin T_n (i) \  and \  \forall j^{''} \notin T_m (j)
\]

\noindent
By substitution we get,

\[
var(\hat{\rho}_{nm}^U ) = nm\times
\begingroup
\renewcommand*{\arraystretch}{2.0}
\begin{bmatrix}
     \frac{1}{4} +\ \big( n-a(n,\theta) \big) \times \frac{1}{12} \\
     +\ \big(a(n,\theta ) - 1 \big) \times E\big( C_{ij-i^{'} j} \mid i^{'} \in T^{-}_{n}(i) \big) \\
     +\ \bigg( m - \frac{a ( n , \theta )\times m}{n}\bigg) \times \frac{1}{12} \\
     +\ \bigg(\frac{a(n,\theta) \times m}{n} -1\bigg) \times E(C_{ij-ij^{'}} \mid j^{'} \in T^{-}_{m}(j)) \\
     +\ \big(n-a(n,\theta)\big) \times \frac{}{} \bigg(m - \frac{a(n, \theta)\times m}{n}\bigg)\times 0 \\
     +\ \bigg(\frac{a(n,\theta) \times m}{n} - 1\bigg) \times  \big(a(n,\theta) - 1\big) \times E( C_{ij-i^{'}j^{'}} \mid j^{'} \in T^{-}_{m}(j) , i^{'} \in T^{-}_{n}(i)) \\
     +\ \big(n - a(n, \theta)\big) \times \bigg(\frac{a(n,\theta) \times m}{n} -1\bigg) \times E\big( C_{ij-i^{''}j^{'}} \mid j^{'} \in T^{-}_{m}(j) , i^{''} \notin T_{n}(i) \big) \\
     +\ \big(a(n, \theta) -1 \big) \times \bigg( m -\frac{a(n, \theta)\times m}{n} \bigg) \times E\big( C_{ij-i^{'}j^{''}} \mid j^{''} \notin T_{m}(j) , i^{'} \in T^{-}_{n}(i) \big)
  \end{bmatrix}
\endgroup
\]

Let
\[
a = a(n,\theta)
\]
\[
C_1= E(C_{ij-i^{'}j^{''}} \mid i^{'}\in T^{-}_n(i) ,\   j^{''} \notin T_m(j) )= E(C_{ij-i^{''}j^{'}} \mid i^{''} \notin T_n(i),\ j^{'} \in T^{-}_m(j))
\]
\[
C_2 = E(C_{ij-i^{'}j^{'}} \mid j^{'} \in T^{-}_{m}(j), i^{'} \in T^{-}_{n}(i) )
\]
\[
C_3 = E(C_{ij-i^{'}j} \mid i^{'} \in T^{-}_{n}(i) ) = E( C_{ij-ij^{'}} \mid j^{'} \in T^{-}_{m}(j))
\]

By substitution and simplification, it follows that:

\[
var(\hat{\rho}_{nm}^U ) = nm\times
\begingroup
\renewcommand*{\arraystretch}{2.0}
\begin{bmatrix}
     \frac{1}{4} +\ \big( n -a -1 \big) \times \frac{1}{12} \\
     +\ a \times C_3 \\
     +\ \big( m - \frac{a\times m}{n} -1 \big) \times \frac{1}{12} \\
     +\ \frac{a \times m}{n} \times C_3 \\
     +\ \big(n -a -1\big) \times \frac{}{} \big(m - \frac{a \times m}{n} -1\big)\times 0 \\
     +\ \frac{a \times m}{n} \times  a \times C_2 \\
     +\ \big(n - a -1\big) \times \frac{a \times m}{n} \times C_1 \\
     +\ a \times \big( m -\frac{a\times m}{n} -1 \big) \times C_1
  \end{bmatrix}
\endgroup
\]

\[
\begin{aligned}
var(\hat{\rho}_{nm}^U )= \frac{1}{4} 
+ \frac{n+m-2}{12} 
- \frac{a +\frac{a \times m}{n}}{12}
+ C_3 \times \bigg( a + \frac{a \times m}{n} \bigg) \\
+ C_2 \times \bigg( \frac{a^2 \times m}{n} \bigg)
+ C_1 \times \bigg( 2am -a - \frac{am}{n} - \frac{2ma^2}{n}\bigg)
\end{aligned}
\]

\[
= \frac{1+n+m}{12} 
+ m \times a \times \Bigg[ 
	\bigg( C_3 -\frac{1}{12} \bigg)  \times \bigg( \frac{1}{m} +  \frac{1}{n} \bigg) 
    + C_2 \times \bigg( \frac{a}{n} \bigg)
    + C_1 \times \bigg( 2 - \frac{1}{m} - \frac{1}{n} - \frac{2a}{n} \bigg)
\Bigg]
\]

\noindent
Within the squared brackets, the terms $1/n$ and $1/m$ should vanish faster than the other terms as $n$ and $m$ grow large, and thus, the result can be approximated by:

\[
var(\hat{\rho}_{nm}^U ) \approx \frac{1+n+m}{12} 
 + m \times a \times \Bigg[
     \big(C_2 - 2C_1 \big) \times \bigg( \frac{a}{n} \bigg)
    + 2C_1
\Bigg]
\]

\[
\approx \frac{1+ n+m \times (1+K(n, \theta ))}{12}
\]

\noindent
where 

\[
K(n,\theta )=12 \times a(n, \theta) \times \Bigg[
     \big(C_2 - 2C_1 \big) \times \bigg( \frac{a(n, \theta)}{n} \bigg)
    + 2C_1
\Bigg]
\]

\noindent
Finally, the original statistic $\rho_{nm}^U$ is a simple function of $\hat{\rho}_{nm}^U$
\[
\rho_{nm}^U = 2 \times \hat{\rho}_{nm}^U - 1
\]

And thus,

\[
Var(\rho_{nm}^U ) \approx \frac{1+ n + m \times (1+K(n, \theta))}{3 \times n \times m}
\]
\\
\noindent
Based on the empirical results given in section~\ref{results_sec}, $K(n,\theta )$ can be well extrapolated using a small constant value of 0.5, at least for the proposed classifier. This should not be a surprising result because:

\begin{itemize}
\item
The correlation effect is merely a result of a stochastic localized clustering or a stochastic large spacing within subsets of adjacent observed examples in the 2D rank space under independence.
\item $a(n, \theta)$ is expected to have an inverse relationship with $C_1$ and $C_2$. In that, the larger the distance between two examples, the smaller is the effect of correlation in the classifier output scores.
\item As a result, as $n$ grows large, for any given non-infinitesimal positive $C_1$ and $C_2$, $a(n, \theta)$ is not expected to grow larger with $n$, especially if the classifier complexity is allowed to grow, even if very slowly. In other words,
\[
\lim_{n\to\infty} \frac{a\big(n, \theta|C_1  > 0, C_2 > 0\big)}{n} = 0
\]
\end{itemize}

On the other hand, figure~\ref{fig:null_dist_comparison} shows the distribution of $\rho_{nm}^U$ to be not symmetric around zero. This is likely due to $K(n, \theta )$ being weakly dependent on the value of $\rho_{nm}^U$. This is a minor issue since we care the most about the right most part of the distribution where useful p-values ( > 0.5) need to be approximated.

The derivation given here is by no means a complete proof, but rather a way to make a theoretical justification of the empirical results. More accurate theoretical distribution remains an open problem. 

\subsection{Part C: Software:}

Source code to use the proposed methods or reproduce the shown results is made available: https://github.com/ramimahdi/robust-nonparametric-correlation-based-on-decision-trees

\bibliographystyle{chicago}
\bibliography{main}

\end{document}